\documentclass[cameraready]{Interspeech}

\usepackage{comment}
\usepackage{algorithm}
\usepackage{algpseudocode}
\usepackage{amsmath}
\usepackage{multirow}

\newcommand{\x}{\mathbf{x}}
\newcommand{\m}{\mathbf{m}}
\newcommand{\y}{\mathbf{y}}
\newcommand{\xstego}{\mathbf{\tilde{x}}}
\newcommand{\xspoof}{\mathbf{x_{\mathrm{sp}}}}
\newcommand{\xstegospoof}{\mathbf{\tilde{x}_{\mathrm{sp}}}}
\newcommand{\devdtw}{s_{\mathrm{DTW}}}
\newcommand{\seconds}[1]{#1\,\mathrm{s}}
\newcommand{\khz}[1]{#1\,\mathrm{kHz}}
\newcommand{\kbps}[1]{#1\,\mathrm{kbps}}
\newcommand{\R}{\mathbb{R}} 
\newcommand{\DTW}{\mathrm{DTW}}
\newcommand{\defvspace}{-0.24cm}

\makeatletter
\def\bstctlcite{\@ifnextchar[{\@bstctlcite}{\@bstctlcite[@auxout]}}
\def\@bstctlcite[#1]#2{\@bsphack
  \@for\@citeb:=#2\do{%
    \edef\@citeb{\expandafter\@firstofone\@citeb}%
    \if@filesw\immediate\write\csname #1\endcsname{\string\citation{\@citeb}}\fi}%
  \@esphack}
\makeatother

\title{A Training-Free Proactive Defense Against Partial Speech Manipulation \\ via Self-Embedding Steganography}

\author[affiliation={},orcid=0000-0003-2235-8655]{Yigitcan}{Özer}
\author[affiliation={},orcid=0000-0003-1003-6093]{Zhe}{Zhang}
\author[affiliation={},orcid=0000-0003-3956-0112]{Wanying}{Ge}
\author[affiliation={},orcid=0000-0001-8246-0606]{Xin}{Wang}
\author[affiliation={},orcid=0000-0003-2752-3955]{Junichi}{Yamagishi}
\address{
    National Institute of Informatics, Tokyo, Japan}

\email{\{yiitozer, zhe, gewanying, wangxin, jyamagis\}@nii.ac.jp}


\begin{document}
\bstctlcite{BSTcontrol}
\maketitle

\begin{abstract}
Partial deepfake speech, where only limited segments of an utterance are synthesized or manipulated, poses a significant challenge to existing deepfake detection systems.
As the proportion of spoofed regions decreases, passive detectors become increasingly unreliable, and accurate detection and restoration remain challenging. 
In this paper, we revisit audio steganography from a new perspective and propose its use as a proactive defense against partially deepfaked audio. 
In particular, we consider a self-embedding strategy in which a clean speech signal embeds a compressed representation of itself, enabling post-hoc extraction of reference content.
We demonstrate how existing audio steganography methods can be repurposed to support detection of partial deepfakes through codec-based restoration. 
Experiments on a benchmark dataset show that the proposed approach complements passive defenses.
Remarkably, the proposed method operates without any training, providing a robust and data-efficient alternative for partial deepfake detection.
\end{abstract}

\section{Introduction}
Recent advances in text-to-speech (TTS) and voice conversion (VC) technologies have enabled highly realistic deepfake audio\,\cite{PatelEtAl23_DeepfakeSurvey_Access, BhagtaniEtAl24_DeepfakeSpeechDetectability_IHMMSec}. 
While fully synthesized speech can often be detected reliably\,\cite{LiAP25_SpeechDeepfakeSurvey_ACMCS}, 
real-world attacks increasingly involve \emph{partial} manipulation, where only short segments of an otherwise authentic utterance are replaced\,\cite{ZhangEtAl21_PartialSpoofingDetection_Interspeech}. 
Such localized manipulations substantially reduce the reliability of passive detection systems\,\cite{HeEtAl25_ManipulatedRegionsSurvey_arXiv}, 
which struggle especially when spoofed regions are brief or sparse. 
Moreover, identifying manipulated segments and recovering the original content remains difficult under this setting, 
limiting the practical utility of existing countermeasures\,\cite{WangYamagishi21_SpoofingCountermeas_Interspeech}.
To address these challenges, recent studies have investigated partial deepfake detection and localization\,\cite{CaiEtAl22_ContentDrivenAVDeepfake_DICTA, CaiEtAl23_GlitchInTheMatrix_CVIU, KhanEtAl24_FrameToUtterance_ICASSP, XieEtAl24_TemporaryDeepfake_ICASSP, KleinEtAl25_Pindrop_ACMMM}. 
Furthermore, benchmark datasets such as HalfTruth\,\cite{YiEtAl21_HalfTruth_Interspeech}, PartialSpoof\,\cite{ZhangEtAl23_PartialSpoof_TASLP}, PartialEdit\,\cite{ZhangEtAl25_PartialEdit_Interspeech}, LlamaPartialSpoof\,\cite{LuongEtAl25_LlamaPartialSpoof_ICASSP}, and AV-Deepfake1M\,\cite{CaiEtAl24_AVDeepfake1M_ACMMM} have facilitated research in partial deepfake detection.

In parallel, audio steganography has been widely studied for secure communication and data hiding\,\cite{DjebbarAyad12_AudioSteganographySurvey_EURASIP}. 
Beyond covert communication, steganography also provides a natural mechanism for \emph{proactive defense}, as auxiliary information can be embedded into signals prior to distribution and later extracted for verification and recovery\,\cite{WangEtAl23_DataHidingSurvey_TCSS}. 
Recent neural methods further improve fidelity and capacity, enabling high-quality speech hiding and recovery\,\mbox{\,\cite{KreukEtAl20_HideAndSpeak_Interspeech, CuiEtAl20_ImageIntoAudioSteganography_ICASSP, KongZhang20_AdversarialAudio_Interspeech, GeletaEtAl22_PixInWav_ICASSP, ChangEchizen25_SteganographyMultimodalAI_SciRep, ZhangEtAl25_HiFiStego_TASLP}}. 
Relatedly, several audio watermarking systems, e.g.,\,AudioSeal\,\cite{SanRomanEtAl24_AudioSeal_ICML}, also enable provenance verification and tampering localization. 
However, these approaches primarily focus on authorship and attribution, and do not directly support fine-grained content recovery under partial manipulation.

\begin{figure}[t]
  \centering
  \includegraphics[width=\linewidth]{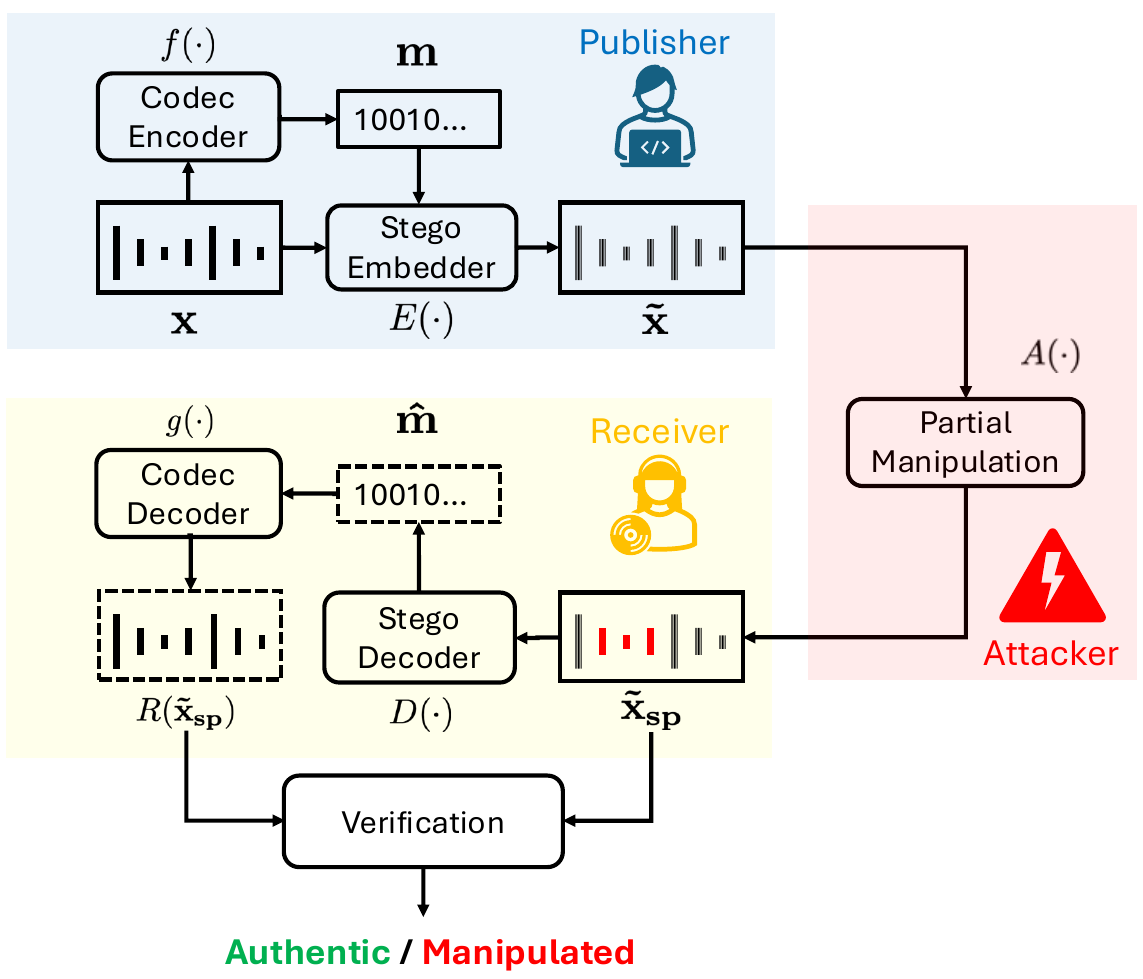}
  \caption{Overview of the proposed self-embedding-based audio steganography framework against partial deepfake manipulation. Discrepancies between the received spoofed signal $\xstegospoof$ and its reconstruction $R(\xstegospoof)$ enable detection while restoration is ensured through neural codec-based reconstruction.}
  \label{fig:mathematical_notation}
  \vspace{-0.48cm}
\end{figure}

In this work, we build upon this perspective and investigate a proactive defense paradigm grounded in audio steganography.
Rather than attempting to identify synthesis artifacts after a manipulation has already occurred, we embed auxiliary information into the signal prior to distribution.
In particular, we adopt a \emph{self-embedding} strategy (see Figure~\ref{fig:mathematical_notation}), in which the embedded payload consists of a compact representation of the carrier signal itself.
Self-embedding steganography has previously been explored in related domains, including video authentication\,\cite{YogarajahEtAl11_VideoAuthentication_IMVIP} and digital image data hiding and reconstruction\,\cite{CheddadEtAl09_SelfEmbedding_SignalProcessing, RakhmawatiEtAl20_BlindSelfEmbedding_IJIES, WangEtAl24_SelfEmbedding_NonlinearDynamics}.
However, to the best of our knowledge, this paper constitutes the first systematic investigation of self-embedding as a proactive defense mechanism in the audio domain.
By embedding a self-referential description of the signal, obtained via a compact neural speech codec representation, our approach enables a direct comparison between the received waveform and its self-reconstruction, thereby facilitating the detection of manipulated regions.

As our main contribution, we propose a temporally repetitive adaptation of the classical least significant bit (LSB) scheme\,\cite{BenderEtAl96_DataHiding_IBM, FranzEtAl96_ComputerSteganography_IH} to embed a neural codec-derived representation of the carrier signal. 
Conceptually, the proposed method is related to the deep learning-based speech hiding method WavInWav\,\cite{FanEtAl25_WavInWav_arXiv}. 
Different from this approach, we adopt a lightweight, training-free strategy that enables robust self-recovery under partial manipulation without requiring large-scale model training or specialized architectures.
Through systematic experiments on a subset of the AV-Deepfake1M dataset\,\cite{CaiEtAl24_AVDeepfake1M_ACMMM}, we demonstrate that the proposed method complements existing passive defenses and outperforms open-source baselines under partial deepfake attacks.

\section{Self-Embedding Steganography \newline Against Partial Speech Manipulation}
\label{sec:proposed_method}
This section formalizes the proposed self-embedding framework for partial speech manipulation, describes the temporally repetitive LSB embedding scheme, analyzes its embedding capacity and repetition behavior, and introduces the computation of the self-reconstruction-based detection score.

\subsection{Problem Formulation}
\label{sec:problem_formulation}
Let $\x \in \R^T$ denote a clean discrete-time speech signal of length $T$.
In general audio steganography, an embedder \mbox{$E : \R^T \times \R^M \rightarrow \R^T$} hides an arbitrary message $\m \in \R^M$ in the carrier signal $\x$, producing a stego signal \mbox{$\xstego = E(\x, \m)$},  which is intended to remain perceptually indistinguishable from the original carrier. 
A corresponding stego decoder \mbox{$D : \R^T \rightarrow \R^M$} attempts to recover the embedded message, resulting in \mbox{$\hat{\m} = D(\xstego)$}.

As depicted in Figure~\ref{fig:mathematical_notation}, this work departs from generic steganography by using a non-arbitrary message $\m$. 
We adopt the neural speech codec SNAC\,\cite{SiuzdakEtAl24_SNAC_NeurIPSWorkshop} with encoder \mbox{$f : \R^T \rightarrow \R^M$}, which maps the carrier to a compact latent representation \mbox{$\m = f(\x)$}.
Under this self-embedding setting, the steganographic embedder hides the codec representation of the signal itself, producing \mbox{$\xstego=E(\x, \m)$}.
Decoding the stego signal yields an estimate  \mbox{$\hat{\m} = D(\xstego)$}.
A synthesis function  \mbox{$g: \R^M \rightarrow \R^T$} is then applied to the recovered message $\hat{\m}$ to obtain a codec-based reconstruction of the waveform.

To analyze the robustness of self-embedding under partial deepfake manipulations, we introduce a spoofing operator $A(\cdot)$ that replaces single or multiple localized time intervals of a signal. 
Let \mbox{$\{\mathcal{I}_k\}_{k=1}^K$} denote a collection of word-level intervals, with each interval given by  \mbox{$\mathcal{I}_k = [t_{\mathrm{start}}^{(k)}, t_{\mathrm{end}}^{(k)}]$}, where $K$ denotes the number of intervals.
For each interval, a corresponding manipulated segment is inserted.
Importantly, the proposed framework makes no assumptions about the internal mechanism of the spoofing operator.
In contrast to passive deepfake detectors that are often trained for specific synthesis artifacts, our formulation treats $A(\cdot)$ as a generic replacement operator.
The manipulated segment may be generated by a TTS or VC model, or obtained through simple signal processing operations such as waveform splicing and segment-level temporal replacement from arbitrary audio recordings.

Building on these definitions, we obtain the following variants of the utterance that serve as the basis for our analysis.  
The \emph{real signal} is the clean carrier $\x$.  
Embedding the self-referential message yields the \emph{stego signal} $\xstego$.  
Applying the partial spoofing operator $A(\cdot)$ to the clean signal produces the \emph{spoofed signal} $\xspoof$\footnote{The $\xspoof$ is not protected by the proposed method; it is used in the experiment to test the performance of passive deepfake detectors (\S\,\ref{sec:baseline}).}, while applying the same operator to the stego signal yields the \emph{spoofed stego signal} $\xstegospoof$.

At verification time, let \mbox{$\y$} denote the received signal, where \mbox{$\y=\xstego$} in the absence of manipulation and \mbox{$\y=\xstegospoof$} under partial spoofing.
We define the self-reconstruction operator as \mbox{$R(\y) = g\big(D(\y)\big).$}
Given a dissimilarity measure $d$, the reconstruction mismatch score is defined as \mbox{$s(\y) = d\big(R(\y),\, \y\big).$} 
Detection is then formulated as a binary hypothesis test,
\[
\delta(\y) =
\begin{cases}
0, & s(\y) \le \tau,\\
1, & s(\y) > \tau,
\end{cases}
\]
where \mbox{$\delta(\y)=0$} indicates an \emph{authentic} signal,
\mbox{$\delta(\y)=1$} indicates a \emph{manipulated} signal,
and $\tau$ is a decision threshold.

\subsection{Temporally Repetitive LSB Adaptation}
\label{sec:temporally_repetitive_lsb}
As a proactive defense mechanism, we adopt LSB embedding, which encodes binary information by modifying the least significant bit of each audio sample while introducing only minimal amplitude distortion.
Specifically, we employ an LSB-based scheme with \emph{temporal repetition}.
For the binary message $\m \in \{0,1\}^M$, the embedder integrates the message bits into the LSB of the audio samples repeatedly at fixed temporal offsets.
In particular, the $r^{\mathrm{th}}$ repetition begins at sample index $rP$, where $P$ is the repetition period (in samples) and $P \ge M$ ensures that repetitions do not overlap.  
Given the carrier signal $\x$ of length $T$, the maximum number of repetitions is then
\[
R = 1 + \left\lfloor \frac{T - M}{P} \right\rfloor .
\]
At decoding time, the decoder reads the LSB stream from the stego audio.
If only a single copy is present, the first $M$ bits are returned. 
When multiple repetitions are available, which is the case in our approach, the decoder retrieves all $R$ copies and applies majority voting across repetitions for each bit:
\[
\hat{m}_i =
\begin{cases}
1, & \text{if } \frac{1}{R}\sum_{r=1}^R m_i^{(r)} \ge 0.5,\\
0, & \text{otherwise}.
\end{cases}
\]

\subsection{Capacity and Repetition Analysis}
\label{sec:capacity}
In our experiments, the carrier waveform is sampled at $\khz{16}$, while the self-embedding payload is obtained by resampling to $\khz{24}$ and encoding with the SNAC $\khz{24}$ model at $\kbps{0.98}$\,\cite{SiuzdakEtAl24_SNAC_NeurIPSWorkshop}.
The resulting bitstream is embedded in the $\khz{16}$ carrier via temporally repetitive LSB encoding, and the same resampling is applied when decoding the payload for self-reconstruction.

At $\kbps{0.98}$, a one-second speech signal produces approximately $980$ payload bits.
Including the $64$-bit synchronization preamble and setting $P=M$ yields a frame length of $1044$ bits.
Since LSB embedding uses one bit per sample, a one-second utterance offers $16{,}000$ embedding positions, allowing $\lfloor 16{,}000 / 1044 \rfloor = 15$ non-overlapping frame repetitions.
This redundancy yields error-free payload recovery in all our experiments, both for unmodified stego signals ($\xstego$) and after the partial manipulation attacks ($\xstegospoof$), described in \S\,\ref{sec:dataset_threat_model}, resulting in $100\%$ bit-exact payload reconstruction and successful waveform decoding in all evaluated cases.

\begin{figure}[t]
    \centering
    \includegraphics[width=\linewidth]{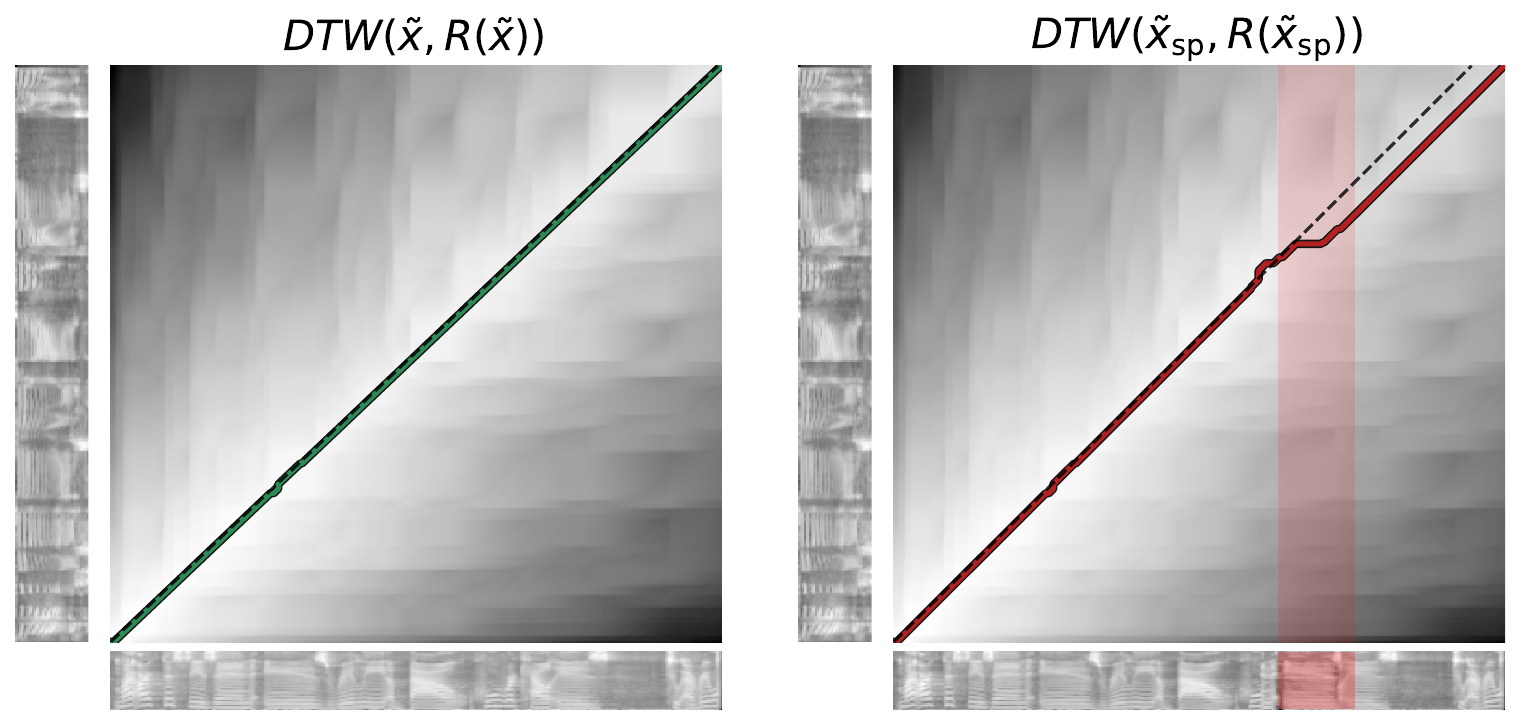}
    \caption{Warping paths computed with dynamic time warping (DTW), \textbf{left:} the stego signal $\xstego$ and the reconstructed audio $R(\xstego)$, without no manipulation, \textbf{right:} the stego signal after manipulation ($\xstegospoof$) and the reconstructed audio ($R(\xstegospoof)$) from the manipulated stego signal.}
    \vspace{\defvspace}
    \label{fig:dtw_manipulation}
\end{figure}

\subsection{Detection Score via Dynamic Time Warping}
\label{sec:eval_method}
For the proposed self-embedding steganography approach, a detection score is computed on the basis of the dynamic time warping (DTW)\,\cite{Muller07_DTW_IRMM} between the received signal $\y$ and its self-reconstruction $R(\y)=g(D(\y))$ (see \S\,\ref{sec:problem_formulation}).
DTW provides a mechanism for aligning two time sequences under possible local temporal distortions. 
This is particularly suitable in our setting, as partial speech manipulations introduce temporal inconsistencies between $\y$ and $R(\y)$.
Consequently, by allowing non-linear temporal alignment, DTW isolates structural mismatches beyond trivial timing offsets.

As illustrated in Figure\,\ref{fig:dtw_manipulation}, we apply the DTW to normalized mel-log spectrogram features using cosine distance as the local dissimilarity measure.
Here, the accumulated cost matrix is shown in grayscale, and the optimal warping path is overlaid in black, while the manipulated segment is highlighted in red.
In the absence of manipulation, the warping path closely follows the diagonal, reflecting near one-to-one temporal correspondence. In contrast, partial manipulations result in higher accumulated costs and localized deviations from the diagonal.

Let $\pi^*$ denote the optimal warping path obtained by $\DTW(R(\y),\y)$.
We define the per-utterance score as
\begin{equation*}
\devdtw(\y) = \frac{1}{|\pi^*|} \sum_{t \in \pi^*} c_t
\end{equation*}
where $c_t$ denotes the local cosine distance between two frames of the mel-log spectrum along the optimal warping path.
Overall, the scalar $\devdtw(\y)$ quantifies the average alignment mismatch between self-reconstruction and the received signal, with higher values indicating a higher chance of being manipulated.

\section{Experiments}
\label{sec:experiments}
This section describes the experimental setup and reports quantitative results for evaluating the proposed self-embedding-based defense against partial deepfake attacks.

\subsection{Dataset and Threat Model}
\label{sec:dataset_threat_model}
The evaluation is conducted on a subset of the \emph{validation split}\footnote{This choice is motivated by the fact that labels and metadata are available only for the training and validation sets, whereas the test set is not publicly annotated.} of the AV-Deepfake1M Dataset\,\cite{CaiEtAl24_AVDeepfake1M_ACMMM}, whose first release is derived from the VoxCeleb2 corpus\,\cite{ChungNZ18_VoxCeleb2DeepSpeaker_Interspeech}. 
Although the database was originally designed for audio--visual deepfake research, our study focuses exclusively on the audio modality. 
Furthermore, we only use the \emph{real} recordings for the experiments.

Although we impose no restrictions on the spoofing operator $A(\cdot)$ (\S\,\ref{sec:problem_formulation}), and the proposed method can be applied to broader cases.
For the proof‑of‑concept experiment in this paper, we assume that the attacker manipulates a part of the real recording. 
%
In particular, the attacker replaces one or two target words with re‑synthesized segments drawn from other utterances of the same speaker, thereby preserving speaker identity while altering the linguistic content. 
In our experiments, the re‑synthesis is conducted using publicly available pre‑trained vocoders or waveform reconstruction algorithms: GriffinLim\,\cite{GriffinL84_SpecgramInversion_TASSP}, HiFiGAN\,\cite{KongKB20_HiFiGAN_SpeechSynthesis_NeurIPS}, HNSincNSF\,\cite{WangYamagishi19_HNSincNSF_SSW}, HNSincNSFHiFi\,\cite{TomashenkoEtAl22_VoicePrivacy_arXiv}, and WaveGlow\,\cite{PrengerVC19_WaveGlow_ICASSP}. 
The attacker is also assumed to use a time‑domain cross‑correlation method when replacing the swapped word\,\cite{ZhangEtAl21_PartialSpoofingDetection_Interspeech}, which helps to reduce the artifacts around the swapped word segments and the chance of being detected.
Note again that the attacker assumed in the experiment is for the proof-of-concept purpose. Testing against diverse attacking methods is left to future work.

\subsection{Evaluation Methodology and Baseline Systems}
\label{sec:baseline}
To assess the detection capability of the proposed self-embedding steganography approach, we quantify the performance using the equal error rate (EER).
The scores $\devdtw(\y)$ of the positive data are computed between the signals without manipulation and their self-reconstruction (i.e.,\,$\DTW(R(\xstego), \xstego)$); those of the negative data are computed between the manipulated signals and the reconstructed versions (i.e.,\,$\DTW(R(\xstegospoof), \xstegospoof)$). 
The decision threshold of EER ($\tau$ in \S\,\ref{sec:problem_formulation}) corresponds to the operating point at which the false positive rate equals the false negative rate.
Lower EER values indicate better discriminative capacity between authentic and manipulated signals.

For comparison, we consider the LAV-DF\,\cite{CaiEtAl22_ContentDrivenAVDeepfake_DICTA} and LAV-DF+\,\cite{CaiEtAl23_GlitchInTheMatrix_CVIU} models, which were originally proposed for audio--visual partial deepfake detection and jointly exploit both modalities.
To the best of our knowledge, these models are among the few partial deepfake detectors with publicly available pretrained checkpoints, enabling reproducible evaluation\footnote{\url{https://github.com/ControlNet/LAV-DF}}.
In our experiments, we operate them in an audio-only setting.
Detection for these two baseline systems relies on the predicted boundary maps produced by the model when the spoofed signal $\xspoof$ is provided as input (see \S\,\ref{sec:problem_formulation}).
A detection score, indicating the chance that the recording is manipulated, is computed as the mean of the boundary map.
The EERs are then computed in the same manner as that for the proposed method.

In addition, we include a pretrained ResNet-based model designed for deepfake word detection~\cite{TranEtAl26_DeepfakeWordDetection_arXiv}. 
Unlike LAV-DF and LAV-DF+, this model operates purely on acoustic features without cross-modal information. 
Including this model allows us to assess whether a strong audio-only detector can identify the word-level attacks considered in our setting. 
Note that the model, by design, produces frame-level detection logits, averages the logits corresponding to each word segment into a single value, and makes a real/fake decision for each word.
For the experimental comparison, wherein an utterance-level score is needed, we compute the mean of the frame-level logits and use it as the utterance-level score.
The EER is then computed in the same manner as for the other methods.

\begin{table}[t]
\centering
\caption{Detection performance under the word-swapping attack, reported as EER (\%). The pooled result is followed by results for each re-synthesis model. Lower EER values indicate better detection performance.}
\resizebox{\linewidth}{!}{%
\begin{tabular}{lcccc}
\toprule 
 & \textbf{Ours} 
 & \textbf{LAV-DF\,\cite{CaiEtAl22_ContentDrivenAVDeepfake_DICTA}}
 & \textbf{LAV-DF+\,\cite{CaiEtAl23_GlitchInTheMatrix_CVIU}}
 & \textbf{ResNet\,\cite{TranEtAl26_DeepfakeWordDetection_arXiv}} \\

\midrule
\multicolumn{5}{c}{\textit{Single-word swapping}} \\

GriffinLim\,\cite{GriffinL84_SpecgramInversion_TASSP}     & 9.95 & 50.17 & 49.90  & 47.46 \\ 
HiFiGAN\,\cite{KongKB20_HiFiGAN_SpeechSynthesis_NeurIPS}  & 9.38 & 50.04 & 49.90  & 47.52 \\ 
HNSincNSF\,\cite{WangYamagishi19_HNSincNSF_SSW}           & 9.52 & 50.10 & 49.90  & 46.97 \\ 
HNSincNSFHiFi\,\cite{TomashenkoEtAl22_VoicePrivacy_arXiv} & 9.52 & 50.31 & 49.90  & 47.52 \\ 
WaveGlow\,\cite{PrengerVC19_WaveGlow_ICASSP}              & 8.91 & 50.37 & 50.03  & 43.98 \\ 

\midrule
\multicolumn{5}{c}{\textit{Two-word swapping}} \\

GriffinLim           & 4.97 & 50.07 & 48.34 &  45.89 \\ 
HiFiGAN              & 5.00 & 50.03 & 48.56 &  46.44 \\ 
HNSincNSF            & 5.07 & 49.96 & 48.34 &  45.67 \\ 
HNSincNSFHiFi        & 5.07 & 50.03 & 48.34 &  45.45 \\ 
WaveGlow             & 4.44 & 50.11 & 48.62 &  42.07 \\ 

\bottomrule
\end{tabular}
}
\vspace{\defvspace}
\label{tab:results_word_swap}
\end{table}

\subsection{Detection Performance}
\label{sec:detection_performance}
We evaluate the detection performance under the word-swapping attack, which represents the primary and realistic partial deepfake threat model considered in this study.
Table~\ref{tab:results_word_swap} reports detection performance under the word-swapping attack for the proposed self-embedding method and the LAV-DF\,\cite{CaiEtAl22_ContentDrivenAVDeepfake_DICTA}, LAV-DF+\,\cite{CaiEtAl23_GlitchInTheMatrix_CVIU}, and ResNet\,\cite{TranEtAl26_DeepfakeWordDetection_arXiv} baselines.

For the single-word swapping case, LAV-DF and LAV-DF+ operate close to a random performance level, with EER values consistently around $50\%$ across all vocoders. 
The pretrained ResNet detector achieves slightly lower EERs in the range of $44\%$--$47.5\%$, yet its performance remains near-random, indicating limited sensitivity to localized word-level manipulations. 
In contrast, the proposed method consistently reduces the EER to approximately $8.9\%$--$10.0\%$ across all re-synthesis models.  
This margin shows that self-reconstruction yields more discriminative cues for detecting content-level manipulations.

For the two-word swapping case, a similar trend is observed. 
LAV-DF and LAV-DF+ remain at a random performance level, while the ResNet detector shows moderate improvement, with EER values decreasing to approximately $42.07\%$--$46.44\%$.  
The proposed method further reduces the EER to $4.4\%$--$5.1\%$, indicating increased detection capability as the extent of manipulation increases.  
Replacing two temporally separated words induces a larger discrepancy between the manipulated waveform and the self-reconstruction of the authentic signal, which the proposed method captures effectively, whereas passive detectors remain largely insensitive to such localized alterations.

\begin{figure}[t]
    \centering
    \includegraphics[width=\linewidth]{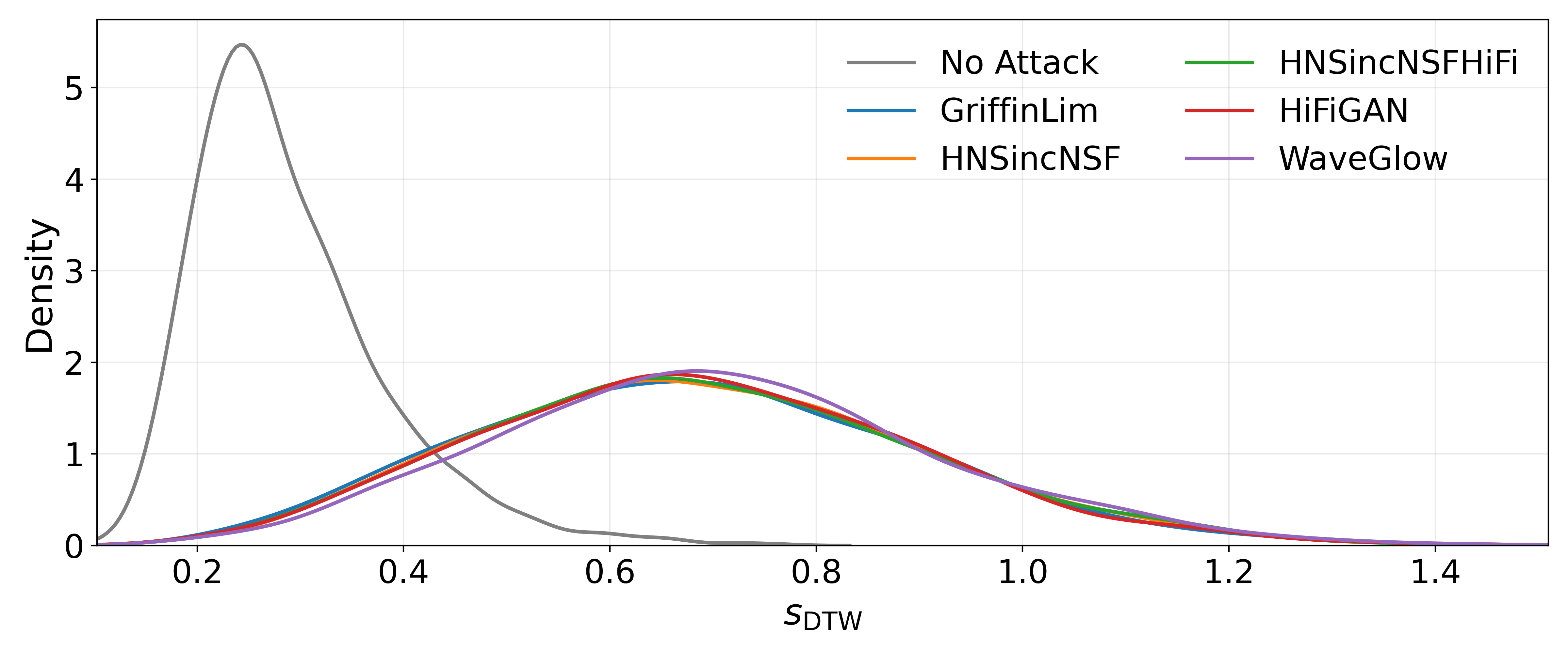}
    \caption{Score distributions computed via DTW-based dissimilarity metric ($\devdtw$) for authentic signals and single-word swap manipulations under various vocoder conditions.}
    \label{fig:dist_attack_no_attack}
\end{figure}

The DTW-based analysis further supports this observation.
Figure~\ref{fig:dist_attack_no_attack} illustrates this behavior by showing the distribution of $\devdtw$ for authentic speech and single-word swap manipulations across all vocoders. 
Authentic samples are tightly concentrated at lower score values, while word-swapping attacks consistently shift the distribution toward higher scores. 
Importantly, this separation persists across different re-synthesis methods, indicating that the proposed self-embedding mechanism captures partial manipulation-induced inconsistencies that remain detectable after vocoder reconstruction.

\begin{figure}[t]
    \centering
    \includegraphics[width=1.01\linewidth]{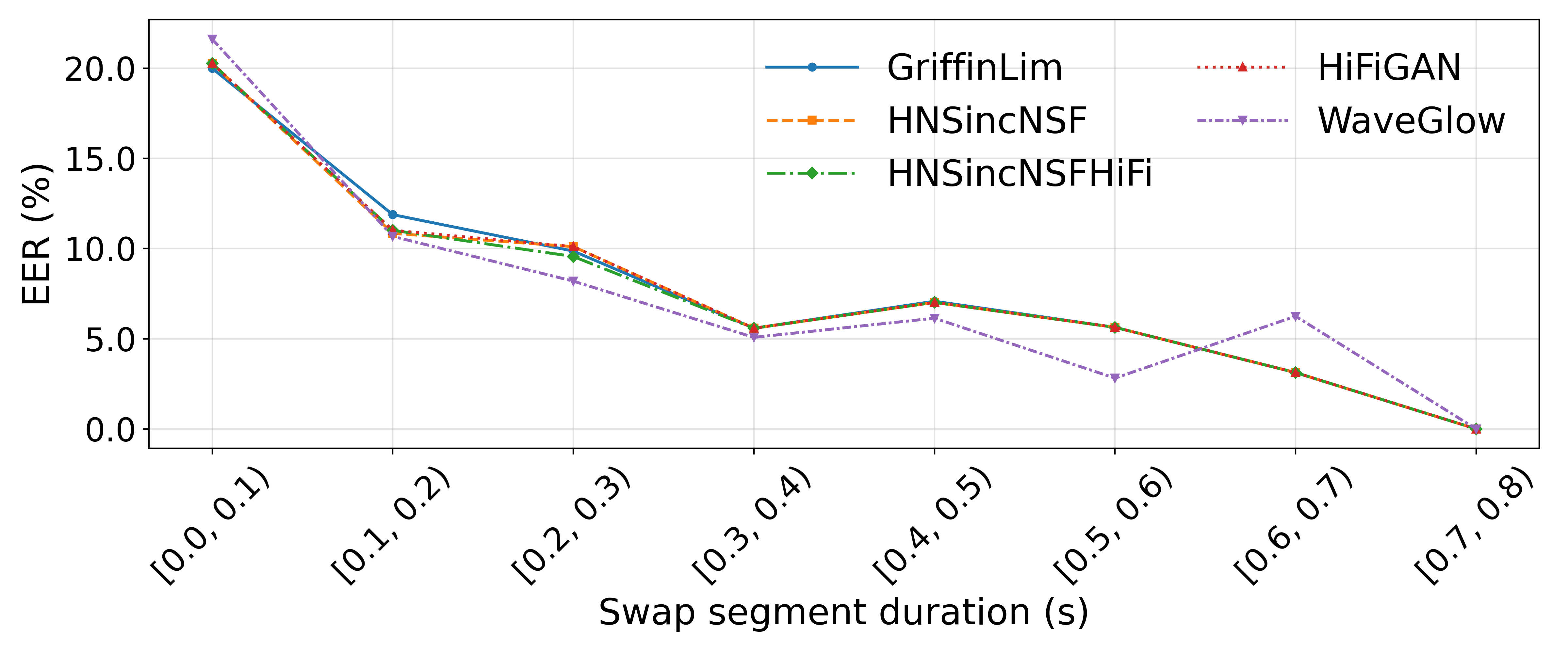}
    \caption{Detection performance as a function of swapped-word duration.
    Each point reports the EER within a 0.1\,s duration bin for different vocoders.}
    \vspace{\defvspace}
    \label{fig:eer_vs_swap_duration}
\end{figure}

Figure~\ref{fig:eer_vs_swap_duration} shows detection performance as a function of the swapped word duration, computed in non-overlapping $\seconds{0.1}$ bins for each vocoder.  
A clear monotonic trend emerges: shorter manipulations are harder to detect, while longer swaps yield progressively lower EERs.  
For very short insertions \mbox{($<\seconds{0.1}$)}, the EER exceeds $20\%$, whereas it drops below $10\%$ once the duration of the swapped segment exceeds $\seconds{0.3}$.  
This pattern aligns with the self-reconstruction-based verification mechanism: longer manipulations lead to a stronger mismatch between $R(\y)$ and $\y$ and consequently lower error rates.

The results reveal a fundamental difference between proactive and passive defenses for partial deepfake detection. 
Passive detectors depend on synthesis artifacts that diminish as the manipulated region becomes shorter, which explains their near-random performance in the word-swapping scenario. 
In contrast, the proposed self-embedding approach evaluates the temporal consistency between the received signal and the self-reconstruction of the authentic waveform, making it capable of detecting even brief semantic or temporal disruptions.
However, when the manipulated segment becomes extremely short, the resulting mismatch becomes weaker, making such manipulations more difficult to detect.

\section{Conclusion}
\label{sec:conclusion}
In this paper, we investigated audio steganography as a proactive defense mechanism against partial manipulations in speech signals.  
By adopting a self-embedding strategy based on temporally repetitive LSB encoding, we showed that a speech signal can reliably carry a compact representation of itself, enabling post-hoc detection of manipulated regions through codec-based restoration.  
Practically, the method is lightweight, training-free, and compatible with existing codecs and embedding schemes, making it readily deployable within current audio processing pipelines.  
Experimental results on a benchmark dataset demonstrated that the proposed approach is effective even when manipulated segments are sparse, without requiring any spoofed training data.  
These findings highlight the complementary roles of proactive and passive defenses for mitigating partial deepfakes.  
Future work will explore more advanced and robust embedding strategies as well as tighter integration with neural codecs to further improve resilience against increasingly sophisticated manipulation attacks.

\section{Acknowledgements}
This work was supported by JSPS MEXT KAKENHI Grant (24H00732). We thank NII visiting researchers Sunan Zou and Yassine El Kheir for their valuable feedback on the article's visualizations.

\section{Generative AI Use Disclosure}
Generative AI tools were used only for language editing and polishing (e.g.,\,grammar, wording, and readability improvements) in this manuscript. No generative AI tool was used to write major parts of the paper, generate scientific claims, or make research decisions. All technical content, analyses, results, and conclusions were produced and verified by the authors, and the authors take full responsibility for the final manuscript.

\bibliographystyle{IEEEtran}
\bibliography{bibliography}

@misc{TranEtAl26_DeepfakeWordDetection_arXiv,
  author       = {Hoan My Tran and Xin Wang and Wanying Ge and Xuechen Liu and Junichi Yamagishi},
  title        = {Deepfake Word Detection by Next-Token Prediction Using Fine-Tuned Whisper},
  howpublished = {arXiv},
  year         = {2026},
  eprint       = {2602.22658},
}

@article{ChangEchizen25_SteganographyMultimodalAI_SciRep,
  author  = {Chia-Chen Chang and Isao Echizen},
  title   = {Steganography Beyond Space-Time with Chain of Multimodal {AI}},
  journal = {Scientific Reports},
  volume  = {15},
  pages   = {12908},
  year    = {2025},
  doi     = {10.1038/s41598-025-97238-2}
}

@misc{FanEtAl25_WavInWav_arXiv,
  author    = {Wei Fan and Kejiang Chen and Xiangkun Wang and Weiming Zhang and Nenghai Yu},
  title     = {{W}av{I}n{W}av: Time{-}Domain Speech Hiding via Invertible Neural Network},
  howpublished = {arXiv},
  year      = {2025},
  eprint    = {2510.02915}
}

@misc{HeEtAl25_ManipulatedRegionsSurvey_arXiv,
  author       = {Jiayi He and Jiangyan Yi and Jianhua Tao and Siding Zeng and Hao Gu},
  title        = {Manipulated Regions Localization for Partially Deepfake Audio: A Survey},
  howpublished = {arXiv},
  year         = {2025},
}

@inproceedings{KleinEtAl25_Pindrop_ACMMM,
  author    = {Nicholas Klein and Hemlata Tak and James Fullwood and Krishna Regmi and Leonidas Spinoulas and Ganesh Sivaraman and Tianxiang Chen and Elie Khoury},
  title     = {Pindrop It! Audio and Visual Deepfake Countermeasures for Robust Detection and Fine{-}Grained Localization},
  booktitle = {Proceedings of the  {ACM} International Conference on Multimedia ({ACMMM})},
  address   = {Dublin, Ireland},
  pages     = {13700--13706},
  year      = {2025},
  doi       = {10.1145/3746027.3761981}
}

@article{LiAP25_SpeechDeepfakeSurvey_ACMCS,
  author  = {Menglu Li and Yasaman Ahmadiadli and Xiao-Ping Zhang},
  title   = {A Survey on Speech Deepfake Detection},
  journal = {{ACM} Computing Surveys},
  volume  = {57},
  number  = {7},
  pages   = {1--38},
  year    = {2025}
}

@inproceedings{LuongEtAl25_LlamaPartialSpoof_ICASSP,
  author    = {Hieu{-}Thi Luong and Haoyang Li and Lin Zhang and Kong Aik Lee and Eng Siong Chng},
  title     = {{L}lama{P}artial{S}poof: An {LLM}{-}Driven Fake Speech Dataset Simulating Disinformation Generation},
  booktitle = {Proceedings of the {IEEE} International Conference on Acoustics, Speech, and Signal Processing ({ICASSP})},
  address   = {Hyderabad, India},
  pages     = {1--5},
  year      = {2025},
  doi       = {10.1109/ICASSP49660.2025.10888070}
}

@article{ZhangEtAl25_HiFiStego_TASLP,
  author    = {Sanfeng Zhang and Baiyu Tian and Yang Gao and Xingyu Liu and Wang Yang},
  title     = {{HIFI}{-}{S}tego: {A} High{-}Fidelity Embedding Audio Steganography Based on Audio Features Decoupling},
  journal   = {IEEE/ACM Transactions on Audio, Speech and Language Processing},
  year      = {2025},
  pages     = {2032--2044},
  doi       = {10.1109/TASLPRO.2025.3570942}
}

@inproceedings{ZhangEtAl25_PartialEdit_Interspeech,
  author    = {You Zhang and Baotong Tian and Lin Zhang and Zhiyao Duan},
  title     = {{PartialEdit}: Identifying Partial Deepfakes in the Era of Neural Speech Editing},
  booktitle = {Proceedings of the Annual Conference of the International Speech Communication Association (Interspeech)},
  address   = {Rotterdam, The Netherlands},
  pages     = {5353--5357},
  year      = {2025}
}

@inproceedings{BhagtaniEtAl24_DeepfakeSpeechDetectability_IHMMSec,
  author    = {Kratika Bhagtani and Amit Kumar Singh Yadav and Paolo Bestagini and Edward J. Delp},
  title     = {Are Recent Deepfake Speech Generators Detectable?},
  booktitle = {Proceedings of the {ACM} Workshop on Information Hiding and Multimedia Security ({IH{\&}MMSec})},
  address   = {Baiona, Spain},
  pages     = {277--282},
  year      = {2024},
  doi       = {10.1145/3658664.3659658}
}

@inproceedings{CaiEtAl24_AVDeepfake1M_ACMMM,
  author    = {Zhixi Cai and Shreya Ghosh and Aman Pankaj Adatia and Munawar Hayat and Abhinav Dhall and Tom Gedeon and Kalin Stefanov},
  title     = {{AV{-}Deepfake1M}: A Large{-}Scale {LLM}{-}Driven Audio{-}Visual Deepfake Dataset},
  booktitle = {Proceedings of the  {ACM} International Conference on Multimedia ({ACMMM})},
  address   = {Melbourne, Australia},
  pages     = {7414--7423},
  year      = {2024},
  doi       = {10.1145/3664647.3680795}
}

@inproceedings{KhanEtAl24_FrameToUtterance_ICASSP,
  author    = {Awais Khan and Khalid Mahmood Malik and Shah Nawaz},
  title     = {Frame-to-Utterance Convergence: A Spectra-Temporal Approach for Unified Spoofing Detection},
  booktitle = {Proceedings of the {IEEE} International Conference on Acoustics, Speech and Signal Processing ({ICASSP})},
  address   = {Seoul, Republic of Korea},
  pages     = {10761--10765},
  year      = {2024},
  doi       = {10.1109/ICASSP48485.2024.10447500}
}

@inproceedings{SanRomanEtAl24_AudioSeal_ICML,
  author    = {Robin San Roman and Pierre Fernandez and Hady Elsahar and Alexandre D{\'{e}}fossez and Teddy Furon and Tuan Tran},
  title     = {Proactive Detection of Voice Cloning with Localized Watermarking},
  booktitle = {Proceedings of the International Conference on Machine Learning ({ICML})},
  year      = {2024},
  address   = {Vienna, Austria}
}

@inproceedings{SiuzdakEtAl24_SNAC_NeurIPSWorkshop,
  author    = {Hubert Siuzdak and Florian Gr{\"o}tschla and Luca A. Lanzend{\"o}rfer},
  title     = {{SNAC}: Multi{-}Scale Neural Audio Codec},
  booktitle = {Audio Imagination: {N}eur{IPS} 2024 Workshop on {AI}{-}Driven Speech, Music, and Sound Generation},
  address   = {Vancouver, Canada},
  year      = {2024}
}

@inproceedings{XieEtAl24_TemporaryDeepfake_ICASSP,
  author    = {Yuankun Xie and Haonan Cheng and Yutian Wang and Long Ye},
  title     = {An Efficient Temporary Deepfake Location Approach Based on Embeddings for Partially Spoofed Audio Detection},
  booktitle = {Proceedings of the {IEEE} International Conference on Acoustics, Speech and Signal Processing ({ICASSP})},
  address   = {Seoul, Republic of Korea},
  pages     = {966--970},
  year      = {2024},
  doi       = {10.1109/ICASSP48485.2024.10448196}
}

@article{WangEtAl24_SelfEmbedding_NonlinearDynamics,
  author  = {L. Wang and S. Banerjee and Y. Cao and J. Mou and B. Sun},
  title   = {A New Self-Embedding Digital Watermarking Encryption Scheme},
  journal = {Nonlinear Dynamics},
  volume  = {112},
  number  = {10},
  pages   = {8637--8652},
  year    = {2024},
  doi     = {10.1007/s11071-024-09521-y}
}

@article{CaiEtAl23_GlitchInTheMatrix_CVIU,
  author  = {Zhixi Cai and Shreya Ghosh and Abhinav Dhall and Tom Gedeon and Kalin Stefanov and Munawar Hayat},
  title   = {Glitch in the Matrix: A Large{-}Scale Benchmark for Content{-}Driven Audio{-}Visual Forgery Detection and Localization},
  journal = {Computer Vision and Image Understanding},
  year    = {2023},
  pages   = {103818},
  doi     = {10.1016/j.cviu.2023.103818}
}

@article{PatelEtAl23_DeepfakeSurvey_Access,
  author  = {Yogesh Patel and Sudeep Tanwar and Rajesh Gupta and Pronaya Bhattacharya and Innocent Ewean Davidson and Royi Nyameko and Srinivas Aluvala and Vrince Vimal},
  title   = {Deepfake Generation and Detection: Case Study and Challenges},
  journal = {{IEEE} Access},
  year    = {2023},
  pages   = {143296--143323},
  doi     = {10.1109/ACCESS.2023.3342107}
}

@article{WangEtAl23_DataHidingSurvey_TCSS,
  author  = {Zihan Wang and Olivia Byrnes and Hu Wang and Ruoxi Sun and Congbo Ma and Huaming Chen and Qi Wu and Minhui Xue},
  title   = {Data Hiding with Deep Learning: A Survey Unifying Digital Watermarking and Steganography},
  journal = {{IEEE} Transactions on Computational Social Systems},
  year    = {2023},
  pages   = {2985--2999}
}

@article{ZhangEtAl23_PartialSpoof_TASLP,
  author  = {Lin Zhang and Xin Wang and Erica Cooper and Nicholas W. D. Evans and Junichi Yamagishi},
  title   = {The {P}artial{S}poof Database and Countermeasures for the Detection of Short Fake Speech Segments Embedded in an Utterance},
  journal = {{IEEE}/{ACM} Transactions on Audio, Speech, and Language Processing},
  year    = {2023},
  pages   = {813--825},
  doi     = {10.1109/TASLP.2022.3233236}
}

@inproceedings{CaiEtAl22_ContentDrivenAVDeepfake_DICTA,
  author    = {Zhixi Cai and Kalin Stefanov and Abhinav Dhall and Munawar Hayat},
  title     = {Do You Really Mean That? {C}ontent{-}Driven Audio{-}Visual Deepfake Dataset and Multimodal Method for Temporal Forgery Localization},
  booktitle = {Proceedings of the International Conference on Digital Image Computing: Techniques and Applications ({DICTA})},
  address   = {Sydney, Australia},
  year      = {2022},
  doi       = {10.1109/DICTA56598.2022.10034605}
}

@inproceedings{GeletaEtAl22_PixInWav_ICASSP,
  author    = {Margarita Geleta and Cristina Punti and Kevin McGuinness and Jordi Pons and Cristian Canton and Xavier Gir{\'o}{-}i{-}Nieto},
  title     = {{P}ix{I}n{W}av: {R}esidual Steganography for Hiding Pixels in Audio},
  booktitle = {Proceedings of the {IEEE} International Conference on Acoustics, Speech, and Signal Processing ({ICASSP})},
  address   = {Singapore},
  pages     = {2485--2489},
  year      = {2022},
  doi       = {10.1109/ICASSP43922.2022.9746191}
}

@misc{TomashenkoEtAl22_VoicePrivacy_arXiv,
  author    = {Natalia Tomashenko and Xin Wang and Xiaoxiao Miao and Hubert Nourtel and Pierre Champion and Massimiliano Todisco and Emmanuel Vincent and Nicholas Evans and Junichi Yamagishi and Jean{-}Fran{\c{c}}ois Bonastre},
  title     = {The {V}oice{P}rivacy 2022 Challenge Evaluation Plan},
  howpublished = {arXiv},
  year      = {2022},
  eprint    = {2203.12468}
}

@inproceedings{WangYamagishi21_SpoofingCountermeas_Interspeech,  
  title     = {A Comparative Study on Recent Neural Spoofing Countermeasures for Synthetic Speech Detection},
  author    = {Xin Wang and Junichi Yamagishi},  
  booktitle = {Proceedings of the Annual Conference of the International Speech Communication Association (Interspeech)},
  doi       = {10.21437/Interspeech.2021-702},  
  year      = {2021},
  address   = {Brno, Czech Republic},
  pages     = {4259--4263}
}

@inproceedings{YiEtAl21_HalfTruth_Interspeech,
  author    = {Jiangyan Yi and Ye Bai and Jianhua Tao and Haoxin Ma and Zhengkun Tian and Chenglong Wang and Tao Wang and Ruibo Fu},
  title     = {Half{-}Truth: A Partially Fake Audio Detection Dataset},
  booktitle = {Proceedings of the Annual Conference of the International Speech Communication Association (Interspeech)},
  address   = {Brno, Czechia},
  pages     = {1654--1658},
  year      = {2021},
  doi       = {10.21437/INTERSPEECH.2021-930}
}

@inproceedings{ZhangEtAl21_PartialSpoofingDetection_Interspeech,
  author    = {Lin Zhang and Xin Wang and Erica Cooper and Junichi Yamagishi and Jose Patino and Nicholas W. D. Evans},
  title     = {An Initial Investigation for Detecting Partially Spoofed Audio},
  booktitle = {Proceedings of the Annual Conference of the International Speech Communication Association (Interspeech)},
  address   = {Brno, Czechia},
  pages     = {4264--4268},
  year      = {2021},
  doi       = {10.21437/INTERSPEECH.2021-738}
}

@inproceedings{CuiEtAl20_ImageIntoAudioSteganography_ICASSP,
  author    = {Wenxue Cui and Shaohui Liu and Feng Jiang and Yongliang Liu and Debin Zhao},
  title     = {Multi{-}Stage Residual Hiding for Image{-}Into{-}Audio Steganography},
  booktitle = {Proceedings of the {IEEE} International Conference on Acoustics, Speech, and Signal Processing ({ICASSP})},
  address   = {Barcelona, Spain},
  pages     = {2832--2836},
  year      = {2020},
  doi       = {10.1109/ICASSP40776.2020.9054033}
}

@inproceedings{KongZhang20_AdversarialAudio_Interspeech,
  author    = {Yehao Kong and Jiliang Zhang},
  title     = {Adversarial Audio: A New Information Hiding Method},
  booktitle = {Proceedings of the Annual Conference of the International Speech Communication Association (Interspeech)},
  address   = {Shanghai, China},
  pages     = {2287--2291},
  year      = {2020}
}

@inproceedings{KongKB20_HiFiGAN_SpeechSynthesis_NeurIPS,
  author    = {Jungil Kong and Jaehyeon Kim and Jaekyoung Bae},
  title     = {{H}i{F}i-{GAN}: Generative Adversarial Networks for Efficient and High Fidelity Speech Synthesis},
  booktitle = {Advances in Neural Information Processing Systems ({NeurIPS})},
  year      = {2020},
  pages     = {17022--17033},
}

@inproceedings{KreukEtAl20_HideAndSpeak_Interspeech, 
  author    = {Felix Kreuk and Yossi Adi and Bhiksha Raj and Rita Singh and Joseph Keshet}, 
  title     = {Hide and {S}peak: {T}owards Deep Neural Networks for Speech Steganography}, 
  booktitle = {Proceedings of the Annual Conference of the International Speech Communication Association (Interspeech)},
  address   = {Shanghai, China}, 
  pages     = {4656--4660}, 
  year      = {2020}, 
  doi       = {10.21437/INTERSPEECH.2020-2380} 
}

@article{RakhmawatiEtAl20_BlindSelfEmbedding_IJIES,
  author  = {Lusia Rakhmawati and Suwadi Suwadi and Wirawan Wirawan},
  title   = {Blind Robust and Self-Embedding Fragile Image Watermarking for Image Authentication and Copyright Protection with Recovery Capability},
  journal = {International Journal of Intelligent Engineering \& Systems},
  volume  = {13},
  number  = {5},
  year    = {2020}
}

@inproceedings{PrengerVC19_WaveGlow_ICASSP,
  author={Ryan Prenger and Raffael Valle and Bryan Catanzaro},
  booktitle={Proceedings of the {IEEE} International Conference on Acoustics, Speech and Signal Processing ({ICASSP})},
  title={{W}ave{G}low: {A} {F}low-based {G}enerative {N}etwork for {S}peech {S}ynthesis},
  year={2019},
  address = {Brighton, UK},
  month = {May},
  pages={3617--3621},
}

@inproceedings{WangYamagishi19_HNSincNSF_SSW,
  author    = {Xin Wang and Junichi Yamagishi},
  title     = {Neural Harmonic-plus-Noise Waveform Model with Trainable Maximum Voice Frequency for Text-to-Speech Synthesis},
  booktitle = {Proceedings of the 10th ISCA Speech Synthesis Workshop ({SSW})},
  year      = {2019},
  pages     = {1--6},
  address   = {Vienna, Austria},
  doi       = {10.21437/SSW.2019-1}
}

@inproceedings{ChungNZ18_VoxCeleb2DeepSpeaker_Interspeech,
  author    = {Joon Son Chung and Arsha Nagrani and Andrew Zisserman},
  title     = {{V}ox{C}eleb2: Deep Speaker Recognition},
  booktitle = {Proceedings of the Annual Conference of the International Speech Communication Association (Interspeech)},
  year      = {2018},
  pages     = {1086--1090},
  address   = {Hyderabad, India},
  doi       = {10.21437/Interspeech.2018-1929}
}

@article{DjebbarAyad12_AudioSteganographySurvey_EURASIP,
  author  = {Fatiha Djebbar and Beghdad Ayad},
  title   = {Comparative Study of Digital Audio Steganography Techniques},
  journal = {{EURASIP} Journal on Audio, Speech, and Music Processing},
  year    = {2012},
  doi     = {10.1186/1687-4722-2012-25}
}

@inproceedings{YogarajahEtAl11_VideoAuthentication_IMVIP,
  author    = {Pratheepan Yogarajah and Joan V. Condell and Kevin Curran and Paul McKevitt},
  title     = {Video Authentication: {A} Self-Embedding Steganography Approach},
  booktitle = {Proceedings of the International Machine Vision and Image Processing Conference ({IMVIP})},
  address   = {United Kingdom},
  year      = {2011},
  pages     = {174--189}
}

@article{CheddadEtAl09_SelfEmbedding_SignalProcessing,
  author  = {Abbas Cheddad and Joan V. Condell and Kevin Curran and Paul McKevitt},
  title   = {A Secure and Improved Self-Embedding Algorithm to Combat Digital Document Forgery},
  journal = {Signal Processing},
  volume  = {89},
  number  = {12},
  pages   = {2324--2332},
  year    = {2009},
  doi     = {10.1016/j.sigpro.2009.02.001}
}

@incollection{Muller07_DTW_IRMM,
  author    = {Meinard M{\"u}ller},
  title     = {Dynamic Time Warping},
  booktitle = {Information Retrieval for Music and Motion},
  publisher = {Springer},
  year      = {2007},
  pages     = {69--84},
  doi       = {10.1007/978-3-540-74048-3_4}
}

@article{BenderEtAl96_DataHiding_IBM,
  author  = {Walter Bender and Daniel Gruhl and Norishige Morimoto and Anthony Lu},
  title   = {Techniques for Data Hiding},
  journal = {IBM Systems Journal},
  volume  = {35},
  number  = {3--4},
  pages   = {313--336},
  year    = {1996},
  doi     = {10.1147/sj.353.0313}
}

@inproceedings{FranzEtAl96_ComputerSteganography_IH,
  author    = {Elke Franz and Anja Jerichow and Steffen M{\"o}ller and Andreas Pfitzmann and Ingo Stierand},
  title     = {Computer-Based Steganography: How It Works and Why Therefore Any Restrictions on Cryptography Are Nonsense, at Best},
  booktitle = {Proceedings of Information Hiding},
  address   = {Cambridge, UK},
  pages     = {7--21},
  year      = {1996}
}

@article{GriffinL84_SpecgramInversion_TASSP,
  author       = {Daniel W. Griffin and Jae S. Lim},
  title        = {Signal estimation from modified short-time {F}ourier transform},
  journal      = {{IEEE} Transactions on Acoustics, Speech, and Signal Processing},
  year         = {1984},
  volume       = {32},
  number       = {2},
  pages        = {236--243}
}

\end{document}